\documentclass[reprint,prd,showpacs]{revtex4-1}

\usepackage[T1]{fontenc}
\usepackage[utf8]{inputenc}
\usepackage[brazil, english]{babel}
\usepackage{amsmath}
\usepackage{amsfonts}
\usepackage{amssymb}
\usepackage{amsthm}
\usepackage{tensor}
\usepackage[amssymb]{SIunits}
\usepackage{mathrsfs}
\usepackage{latexsym}
\usepackage{graphicx}
\usepackage{nicefrac}
\usepackage[unicode]{hyperref}
\usepackage{multirow}
\usepackage{paralist}
\usepackage{SIunits}

\newtheorem{theorem}{Theorem}[section]

\newtheorem{proposition}[theorem]{Proposition}
\newtheorem{corollary}[theorem]{Corollary}

\newcommand{\ud}{\ensuremath{\mathrm{d}}}
\newcommand{\linha}{\ensuremath{^{\prime}}}

\hypersetup{
colorlinks = true,
breaklinks = true
}

\begin{document}

\title{How the expansion of the universe determines the causal structure of McVittie spacetimes}

\author{Alan M. da Silva}
\email{alan.maciel@usp.br}

\author{Michele Fontanini}
\email{fmichele@fma.if.usp.br}

\author{Daniel C. Guariento}
\email{carrasco@fma.if.usp.br}

\affiliation{Instituto de Física, Universidade de São Paulo,\\
Caixa Postal 66.318, 05315-970, São Paulo, Brazil}

\begin{abstract}

We present a theorem which allows one to recognize and classify the asymptotic behavior and causal structure of McVittie metrics for different choices of scale factor, establishing whether a black hole or a pair black-white hole appears in the appropriate limit. Incidentally, the theorem also solves an apparent contradiction  present in the literature over the  causal structure analysis of the McVittie solution. Although the classification we present is not fully complete, we argue that this result covers most if not all physically relevant scenarios.

\end{abstract}


\maketitle

\section{Introduction}

The century-old problem of describing a gravitationally bound system in an expanding universe in the frame-set of general relativity has seen many attempts to find a solution. Despite its apparent simplicity, a full understanding of the mechanisms involved when general and realistic systems are considered has yet to be found. This can be seen, for instance, by taking a look at the vast and often contradictory literature on one of the oldest proposed models, and possibly what could be considered the second simplest scenario after Schwarzschild-de~Sitter, namely the McVittie solution \cite{mcvittie-1933}. 

Our current understanding of the history of the universe, built over an increasing amount of data from accurate measurements such as the cosmic microwave background (CMB) and supernovae distances, tells us that we are most likely living in a homogeneous and isotropic universe which is undergoing a phase of accelerated expansion. The $\Lambda$CDM model, our best fit to observations yet, is far from being a complete model, given that it does not properly include baryons and that we lack a reliable understanding of star formation and feedback \cite{silk-2012,*armijo-2011}. Therefore, it is evident that studying how bound systems feel the expansion of the universe, from collapsing star-forming matter up to galaxy superclusters at recent times, is of vital importance for a better understanding of the cosmos.

The first step in studying bound systems interacting with an expanding background is to consider a related problem: the formation and evolution of black holes in an expanding universe. Some of the simplest metrics which display these features belong to the Kustaanheimo-Qvist class of solutions to Einstein's equations for a comoving shear-free perfect fluid in a spherically symmetric configuration \cite{kustaanheimo-1947,stephani-exact,bolejko-2011}. The oldest and perhaps most famous member of this class is the McVittie solution \cite{mcvittie-1933}. Throughout the years it has been studied in this context, either in its original form (see Refs.~\cite{nolan-1998,*nolan-1999,*nolan-1999a,nandra-2012} and references therein) or in a generalized version with a time-dependent mass \cite{faraoni-2007,*gao-2008,carrera-2010,*carrera-rmp-2010,anderson-2011,guariento-2012} 

In this work we consider the original McVittie solution to Einstein's equations, which together with its physical interpretation has been debated for almost 80 years. Recently, there have been considerable advances towards understanding such a metric; for instance, after a long debate it has been proved \cite{kaloper-2010} that the central object satisfies necessary and sufficient conditions to be characterized as a black hole, provided that the line element asymptotes to the Schwarzschild-de~Sitter metric at temporal infinity. In the same work, it has also been established that the metric possesses a singularity which lies in the past of every causal trajectory, the McVittie big bang. Moreover these results have been shown to be valid for a generalized version of the McVittie metric at least for some time-dependent masses \cite{guariento-2012}.

The causal structure of the McVittie spacetime is one of the crucial points that have been debated in the literature. Its analysis led the authors of Ref.~\cite{kaloper-2010} to conclude that the internal apparent horizon present in the solution does asymptote to a black hole horizon. At the same time, following a similar path and using detailed numerical integrations of the light curves, the authors of Ref.~\cite{lake-2011} found that the inner horizon ends up separating the boundary of the spacetime into two sections, a black hole horizon in the future and a white hole horizon in the past. In this brief work we want to focus on this issue to clarify a peculiar aspect of the McVittie solution. We will show, in fact, that the presence or absence of the white hole part, as discussed in Ref.~\cite{lake-2011}, crucially depends on the choice of the function describing the expansion. Incidentally, this means that the causal structures presented in the works cited above do not need to be considered as contradictory, since they are both possible depending on the particular choices of expansion function made in each work.

The aim of the theorem presented in the following sections is to allow one to recognize the asymptotics, and thus the possible embeddings of the spacetime once the expansion function is chosen. Although we succeeded in describing a wide class of possible solutions, the theorem falls short of giving a prediction for a specific class of expansion functions that decay exponentially in a very peculiar way (as will be explained in detail in Sec.~\ref{subsec:dontknow}) for which a case-by-case study is necessary. On the bright side, all realistic models of expansion nicely fit into the two groups of functions for which the theorem is able to identify the corresponding spacetime causal structure.

After a brief summary of the main characteristics of the McVittie solution in Sec.~\ref{sec:causal}, we perform in Sec.~\ref{sec:rmin} an analysis of the inner apparent horizon, around which the behavior of geodesics defines the asymptotic structure of the spacetime. In Sec.~\ref{sec:main-theorem} we present the main result of this paper, a theorem that allows us to find out whether the inner horizon is an accumulation point for geodesics from both above and below or just on one side. Finally, we conclude in Secs.~\ref{sec:example} and \ref{sec:conclusions} with an instructive example---$\Lambda$CDM---and some remarks.

Throughout the paper we use the convention of representing the derivative with respect to the time coordinate $t$ with a dot, and with respect to the radial coordinate $r$ with a prime.

\section{McVittie causal structure}\label{sec:causal}

The McVittie metric with constant mass parameter $m > 0$ can be written in the form \cite{kaloper-2010,guariento-2012}
\begin{equation} \label{mcvittie}
  \ud s^2 = - \left(R^2 - H^2 r^2 \right) \ud t^2 - \frac{2 H r}{R} \ud r \ud t + \frac{\ud r^2}{R^2} + r^2 \ud \Omega^2 \,,
\end{equation}
where  $R (r) = \sqrt{1 - \nicefrac{2 m}{r}}$, and $H (t) = \nicefrac{\dot{a}}{a}$ is the Hubble parameter. $H$ is assumed to be a smooth function with the following properties:
\begin{subequations}
  \begin{gather}
    \lim_{t \to \infty} H (t) = H_0 \,,\label{Hprop1}\\
    \frac{1}{3 \sqrt{3} m} > H_0 >0 \,,\label{H-no-extreme} \\
    \dot{H} (t) < 0 , \quad t > 0 \,.\label{Hprop2}
  \end{gather} 
\end{subequations}

\begin{figure}[!htp]
  \includegraphics[width=0.45\textwidth]{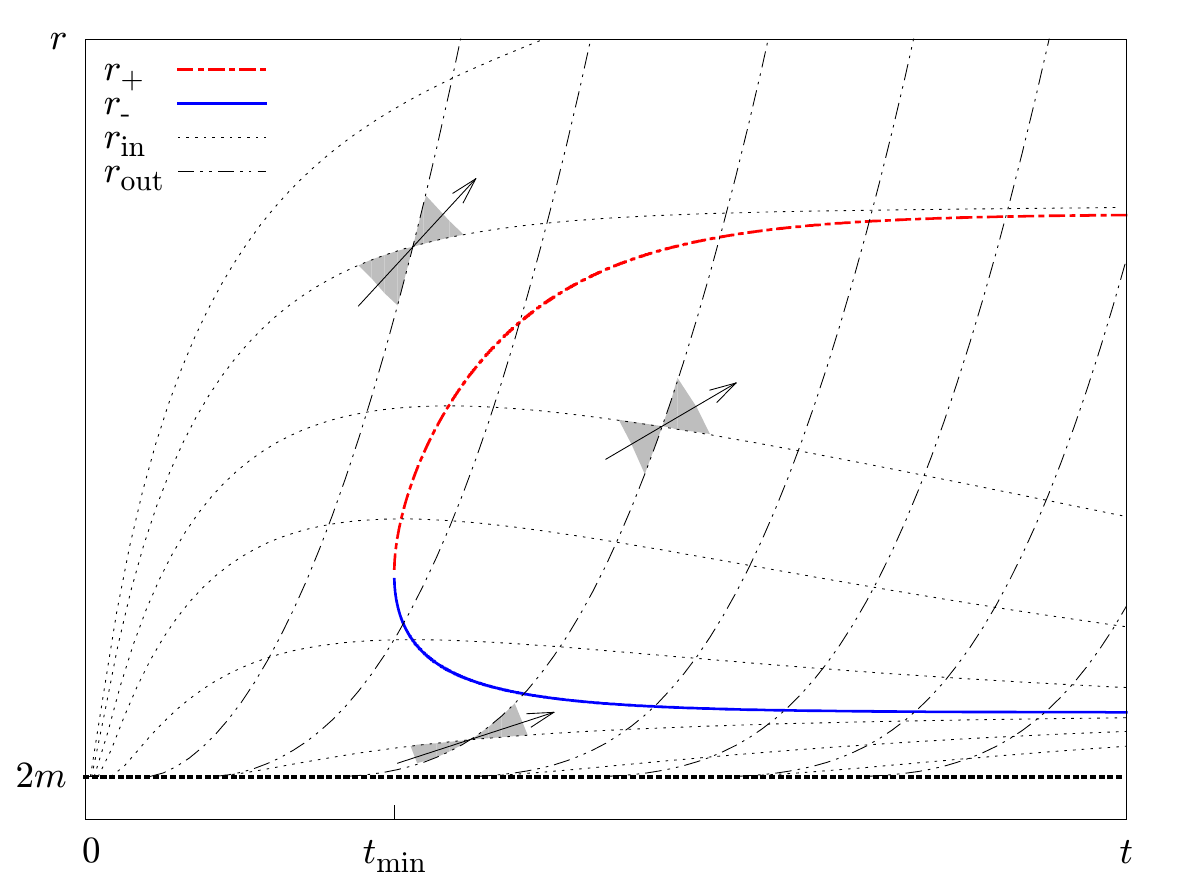}
  \caption{An example of the McVittie apparent horizons plotted in spacetime coordinates. Some ingoing ($r_{\text{in}}$) and outgoing geodesics ($r_{\text{out}}$) are also shown, as well as some light cones, that are represented by the shaded regions, where the arrows indicate the future direction. The singularity at $r=2m$ is represented by a thick dotted line.\hspace*{\fill}}
  \label{horizontes-rt}
\end{figure}

Drawing the Penrose diagram of a fully dynamical metric such as \eqref{mcvittie}, which satisfies certain conditions as described in Ref.~\cite{walker-1970}, reduces to integrating geodesics and is thus mostly a numerical effort. As already pointed out, we will make some remarks on previous works on this topic, and for an easier comparison, we will then adopt the choices made in Ref.~\cite{lake-2011}. Singularities of the metric \eqref{mcvittie} have been studied in Refs.~\cite{lake-2011} and \cite{kaloper-2010}. There, using the fact that all future-oriented null geodesics move away from the singularity in its neighborhood, it has been shown that the spacelike surface defined by $r=2m$ lies to the past of every event of the spacetime covered by our coordinates, and it is thus dubbed the ``McVittie big bang.'' It has also been shown that $t=0$, in general, does not belong to the spacetime. We thus choose the singular surface at $r=2m$ as our reference point, identifying it with a horizontal line in the conformal diagram. As in Ref.~\cite{lake-2011}, every event will then be connected to the McVittie big-bang surface via ingoing (``$-$'') and outgoing (``$+$'') null rays, which are solutions of the geodesic equation
\(
\dot{r} = R (H r \pm R),
\)
as can be seen in Fig.~\ref{horizontes-rt}. We take the function transforming from times to coordinates in the causal diagram to be Eq.~(35) in Ref.~\cite{lake-2011}.

The behavior of the outgoing geodesics is well understood, and their integration does not present difficulties. On the other hand, due to the fact that the apparent horizons of the McVittie metric are antitrapping surfaces, and thus influence the behavior of ingoing geodesics, the latter are to be treated more carefully and present the only source of possible confusion. In particular, the presence of an accumulation point for the ingoing geodesics makes numerical analysis and integration challenging. Therefore, in what follows we focus on such geodesics represented by solutions of the differential equation for the function $r: (t_0, \infty) \to (2 m, \infty)$ \cite{kaloper-2010},
\begin{gather}
  \dot{r} (t)  = R (r) \left[ r H (t)- R (r) \right] \equiv X (t, r) \,, \label{hor-eq}
  \intertext{with initial condition}
  r (t_i) = r_0 > 2 m \,, \label{initialcondition}
\end{gather}
which ensures that the coordinates in \eqref{mcvittie} describe the physical space above the singularity. Note that property \eqref{Hprop2} implies $\dot{X}<0$.

By integrating the ingoing geodesics backward in time, we can define a \emph{first time} $t_0$, the time at which each geodesic leaves the singularity at $r=2m$, taken to be the reference for the conformal diagram. Correspondingly, we will refer to the geodesic starting at $t_0$ as $r_{t_0}(t)$, so that
\[
  \lim_{t \to t_0} r_{t_0} (t) = 2 m \,.
\]

Let us denote $f(t, r) = \frac{X (t, r)}{R (r)}$; the \emph{apparent horizon} $\mathcal{H}$ is the locus of points in the $(t, r)$ plane in which $f(t,r)=0$. In other words, the horizon is the set of points where $X (t, r) = 0$ above the singularity. An example of horizons for the McVittie metric for a specific choice of scale factor is depicted in Fig.~\ref{horizontes-rt}.

It can be easily proved by convexity arguments that, for each fixed value of $t$, $f$ has up to two real roots for positive values of $r$, and the roots always belong to the interval $(2m, \infty)$. Moreover, if $f(\hat{t},r)$ has at least one real root, then $f (t, r)$ has two real roots for every $t > \hat{t}$. It can also be noted that when the two roots are distinct, and thus have multiplicity 1, one has $f\linha(t,r) \neq 0$ at the root.

We call the inner horizon $\mathcal{H}_-$ the locus defined by the smallest real root of $f (t, r)$ for each $t$ for which it exists. Calling $t_{\text{min}}$ the smallest value of $t$ for which $f (t, r)$ admits at least one real root, the inner horizon $\mathcal{H}_-$ is then represented by the curve $(t, r_- (t))$, where $r_-:[t_{\text{min}}, \infty) \to (2 m, \infty)$ is a smooth function. We will call $r_-$ the inner horizon function, or, when there is no room for confusion with the set $\mathcal{H}_-$, just inner horizon for short. All points of $\mathcal{H}_-$ correspond to points in which the geodesic flow is horizontal in the $(t, r)$ plane, as the tangent to the flow is given by the value of $X(t,r)$.

A useful characteristic of the inner horizon function is that it has a negative slope for all times for which it is defined. Explicitly, by taking the gradient of $f(t,r)$ along the curve, the slope reads 
\begin{equation}
\label{negslope}
  \dot{r}_-(t) = \frac{-r_- (t) \dot{H} (t)}{H (t) - R^{\prime}(r_- (t))} \,,
\end{equation}
provided that $f\linha(t, r_- (t))=H (t) - R\linha (r_- (t))  \neq 0$. The numerator of \eqref{negslope} is clearly positive following \eqref{Hprop2}, and it is easy to show that the denominator $f\linha(t, r_- (t)) < 0$. In fact, noticing that $f(t,2m)=2mH(t)>0$, by continuity for any time for which $r_-(t)$ is a simple root, it follows that $f (t, r_- (t) - \delta) > 0$, for $\delta > 0$, ensuring that $f\linha$ is negative on the inner horizon. This also implies that the ingoing geodesics increase monotonically in the inner region $2m<r < r_-$.

\subsection{\texorpdfstring{The first-time projector $\mathbf{\Phi}$}{The first-time projector Φ}}\label{subsec:phi}

To prepare for our main result, we are going to introduce a function which will play a crucial role in the rest of this work, the first-time projector $\Phi$. This function, or better, the image of the inner horizon under it, will contain the information on the causal structure of the spacetime considered. 

Let us call $\mathcal{M}$ the set of events covered by our coordinates $(t, r)$, so that $\mathcal{M} \subset (0, \infty) \times (2 m, \infty)$, and define the application $\Phi: \mathcal{M} \to \mathcal{I}$, where $\mathcal{I} \subset (0, \infty)$, which associates to each element of $\mathcal{M}$ the first time $t_0$ of the ingoing geodesic $r_{t_0}(t)$ which passes by that point. In other words,  $\Phi(\mathcal{M})$ corresponds to the time component of the points in the intersection between the singularity and the image of $\mathcal{M}$ under the ingoing geodesic flow; it is the set of values of $t_0$ one can reach by integrating geodesics, starting at any event in $\mathcal{M}$ back to the singularity.

It is easy to show that $\Phi$ is well defined, in the sense that the image of each coordinate pair $(t, r)$ in its domain is at most one point of the interval (or empty); in fact, it is straightforward to check that the system given by Eqs.~\eqref{hor-eq} and \eqref{initialcondition} satisfies the hypotheses of the Picard--Lindel\"of theorem, which means that the ingoing geodesics at each event above the singularity are unique. Then, the image of an event by the first-time projector $\Phi$, if it exists, is a unique first time $t_0$.

\section{Inner horizon asymptotic behavior}
\label{sec:rmin}

\begin{figure}[!hbp]
  \includegraphics[width=.45\textwidth]{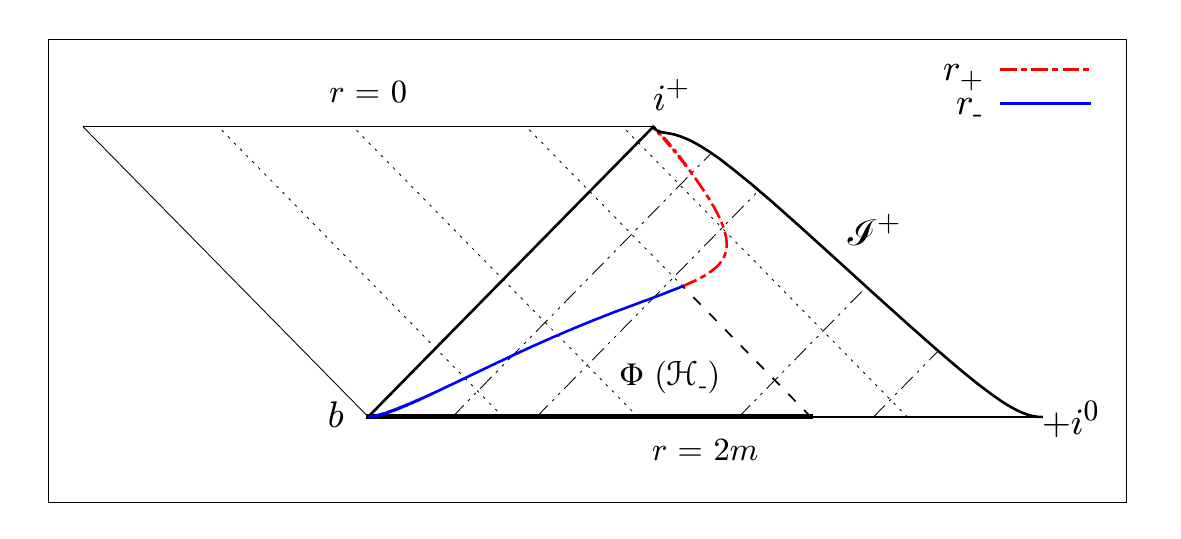}
  \caption{An extension of McVittie spacetime with an unbounded image of the inner horizon over the singularity. Ingoing geodesics are represented by lines inclined 45\degree{} to the left. The left part of the graph shows the appearance of a black hole whose horizon is given by $r_-$ in the limit $t\to \infty$.}
  \label{causal-px}
\end{figure}

\begin{figure}[!hbp]
  \includegraphics[width=0.45\textwidth]{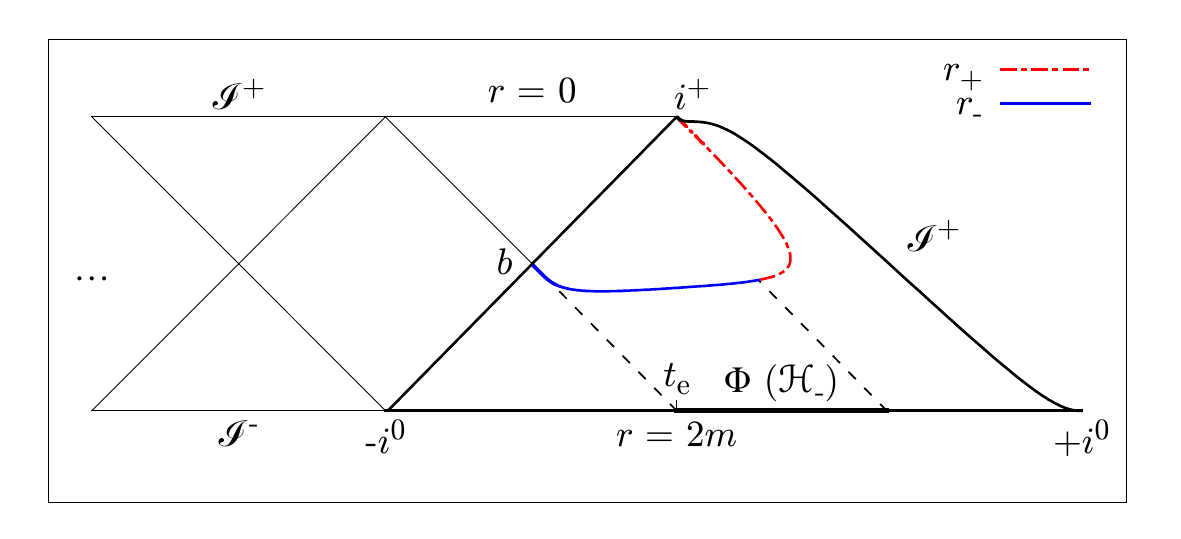}
  \caption{A null geodesically complete extension of a McVittie spacetime characterized by a bounded upper value for $\Phi(\mathcal{H}_-)$, the first-time projector of the inner horizon. Notice that geodesics starting after $t_{\text{e}}$ do not reach the horizon in a finite time. This completion shows the presence of a white hole and a black hole region, separated at the branching surface $b$, as discussed in Ref.~\cite{lake-2011}.\hspace*{\fill}}
  \label{diagrama-bound}
\end{figure}

Why are we interested in the image of the geodesic flow projected on the singularity? The main reason for that can be understood by looking at two cases considered in the literature. In Ref.~\cite{kaloper-2010} one can see that the image of $\mathcal{H}_-$ under the projector $\Phi$ defined in Sec.~\ref{subsec:phi} is unbounded, and the asymptotic analysis shows the presence of a black hole (see Fig.~\ref{causal-px}). On the other hand, the analysis done in Ref.~\cite{lake-2011}, where the projection of the inner horizon is bounded, finds that the spacetime presents a bifurcation two-sphere that separates an asymptotic black hole and a white hole (see Fig.~\ref{diagrama-bound}). We agree with the results presented by both groups, and we will show in what follows that the discrepancies can be explained by assuming that a different choice for the function describing the expansion of spacetime has been made. In fact, we want to show that the connection between the projected flow and the asymptotic behavior is what allows one to distinguish between the two possibilities of inevitably finding a white hole accompanying a black hole or not. So the problem we are facing can be restated as follows: is the image of $\mathcal{H}_-$ under $\Phi$ bounded? A negative answer would imply the presence of a black hole alone, while a positive one would change the structure at infinity and produce a white hole as well.

\begin{figure}[!hbp]
  \includegraphics[width=.45\textwidth]{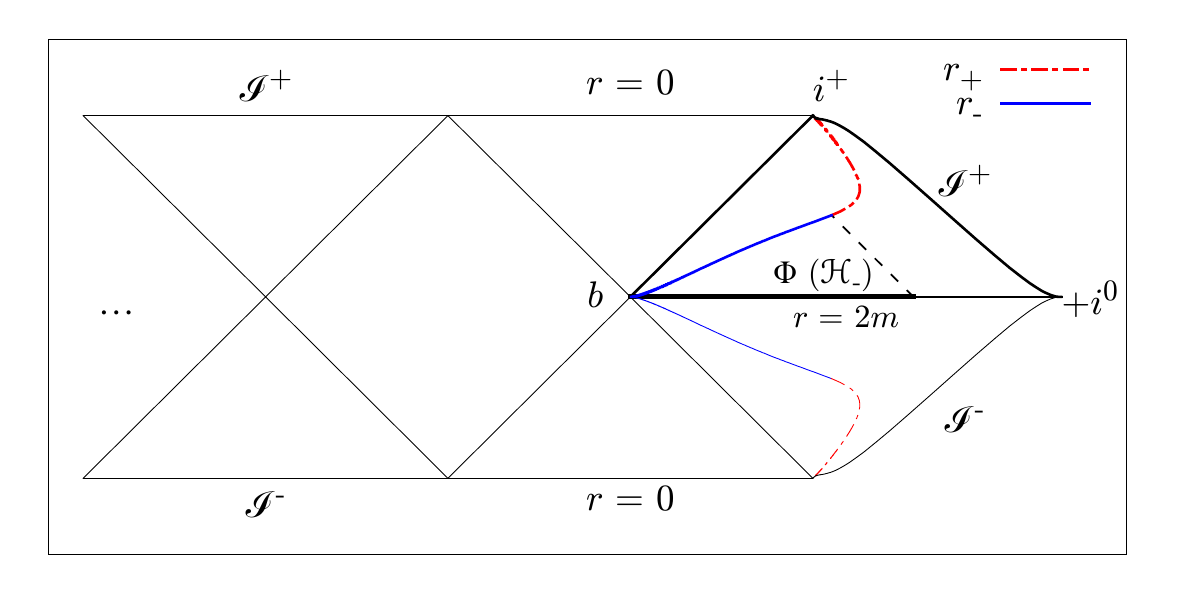}
  \caption{A possible null geodesically complete extension of Fig.~\ref{causal-px}, where a time-reversed McVittie metric with a big crunch also appears \cite{faraoni-2012}.}
  \label{causal-x}
\end{figure}

To answer the question of boundedness, we will need some preliminary results in order to reduce the problem to a simpler one. First, let us define the limiting values at temporal infinity for $r_-(t)$ and $f(t,r)$ as $r_{\infty} \equiv \lim_{t \to \infty} r_-(t)$ and $f_{\infty}(r) \equiv \lim_{t \to \infty} f(t, r)$. Then,
\begin{gather}
  2m < r_{\infty} < r_- (t) \,,\nonumber
\intertext{and}
  f_{\infty} (r_{\infty}) = r_\infty H_0 - R (r_{\infty}) = 0 \,. \label{finfinito}
\end{gather}

We note that, in the case in which the image of the inner horizon by the first-time projector $\Phi$ is bounded, there exists $t_{\text{e}}$ such that, for each $\tau>t_{\text{e}}$, the ingoing geodesic $r_{\tau}(t)$ that leaves the singularity at instant $\tau$ never crosses the inner horizon $r_-(t)$ and tends to it at time infinity. This means that $\Phi(\mathcal{H}_-) \subset (0, t_{\text{e}}]$. 

In what follows we show some consequences of taking as a hypothesis the fact that $\Phi(\mathcal{H}_-)$ is bounded.

\begin{proposition}\label{rlimit}

Let $r_{t_0}(t)$ be an ingoing geodesic.  If $r_{t_0}$ satisfies $r_{t_0}(t) < r_-(t)$ for all $t>t_0$, then
\begin{gather}
r_{t_0} (t) < r_{\infty}, \ \forall\, t > t_0 \,, \label{parte1}
\intertext{and}
\lim_{t \to \infty} r_{t_0}(t) = r_{\infty} \,. \label{parte2}
\end{gather}

\begin{proof}
Consider the sequences  $t_n = t_0 + n \tau$, $r_n = r_{t_0} (t_n)$, $h_n = r_- (t_n)$, $n \in \mathbb{N}$, $\tau > 0$. By hypothesis, and by making use of the monotonic increase of inner geodesics, the two sequences satisfy the following properties:
\begin{subequations}
\begin{gather}\label{rnbounded}
2m \leq r_n < h_n \,,\\
r_n > r_m \iff n>m \,. \label{rncrescente}
\end{gather}
\end{subequations}

To prove \eqref{parte1}, we consider by \emph{reductio ad absurdum} the case $r_n > r_{\infty}$, for some $n$. Then, as $\lim_{n \to \infty} h_n = r_{\infty}$, this would imply $r_n > h_m$, for some $m > n$. But, by property \eqref{rnbounded}, $r_m < h_m$, which implies  $r_n > r_m$ for $n < m$, and contradicts property \eqref{rncrescente}. The case  $r_n = r_{\infty}$ follows by noting that if it holds for some $n$, then $r_{n + 1} > r_n > r_{\infty}$, by property \eqref{rncrescente}, which leads us to the previous case.

Moving on to \eqref{parte2}, note that $r_n$ is monotonic and bounded; therefore, it converges. Let us call $\lim_{n \to \infty} r_n \equiv r^* \leq r_{\infty}$ its limit. Recalling the definition of $X(t,r)$ in \eqref{hor-eq} and the mean value theorem, there exist sequences $\bar{t}_n$ and $\bar{r}_n = r_{t_0}(\bar{t}_n)$ such that 
\begin{gather}
r_n < \bar{r}_n < r_{n + 1} \,,\nonumber\\
r_{n + 1} - r_n = \tau X (\bar{t}_n, \bar{r}_n) \,.\label{usingMVT}
\end{gather}
Since $\lim_{n \to \infty} \bar{r}_n = r^*$, the limit of \eqref{usingMVT} is $0 = \tau R (r^*) f_{\infty} (r^*)$; then for this to hold, it must be that $r^* = r_{\infty}$.
\end{proof}

\end{proposition}

The importance of proposition \ref{rlimit} is clear; in fact, thanks to this proposition, one only has to analyze geodesics crossing $r_\infty$, a fixed surface, eliminating the complication of having to consider the time-varying inner horizon. Moreover, this implies that the $r_\infty$ surface behaves as an accumulation point for the geodesics, as we will see in the following corollary.

\begin{corollary}\label{r-epsilon}

If $r_{t_0} (t)$ is an ingoing geodesic, then for all $\epsilon$ satisfying $0 < \epsilon < r_{\infty} - 2 m$, there exists $\bar{t} > t_0$ such that $r_{t_0} (\bar{t}) = r_{\infty} - \epsilon$.

\end{corollary}

This means that, given enough time, all geodesics either cross $r_{\infty}$ or reach values arbitrarily close to it. By proposition~\ref{rlimit}, every ingoing geodesic that never crosses the inner horizon never reaches $r_{\infty}$. Conversely, if an ingoing geodesic does traverse $r_{\infty}$ in a finite time interval, then it eventually crosses the inner horizon. Then, by studying only the neighborhood of $r_{\infty}$, we may tell if geodesics do or do not cross the horizon.

To complete our set of preliminary results, we will show now that geodesics that start at a later first time remain below (that is, at smaller values of $r$ than) those which start at earlier first times. This can be stated precisely by the following.

\begin{proposition}\label{r1>r2}
Let $r_{t_1} (t)$ and $r_{t_2} (t)$ be two ingoing geodesics. Therefore, if $t_1 < t_2$, then $r_{t_1} > r_{t_2}$.

\begin{proof}
Using the fact that the ingoing geodesic passing by any event $(t, r)$ of the plane is unique for $t > 0$, $r > 2 m$, then we conclude that they cannot cross in the same region. Moreover, as for any $\delta > 0$, $r_{t_0} (t_0 + \delta) > r_{t_0 + \delta} (t_0 + \delta) = 2 m$, then by continuity $r_{t_0} (t) - r_{t_0 + \delta} (t) > 0$ for all $t > t_0 $.
\end{proof}
\end{proposition}

Propositions~\ref{rlimit} and \ref{r1>r2} imply that, if we prove that there exists one ingoing geodesic that never reaches $r_{\infty}$, then $\Phi (\mathcal{H}_-)$ is bounded, as every geodesic which leaves the singularity later will never reach the inner horizon as well. Otherwise, $\Phi (\mathcal{H}_-)$ is unbounded if every ingoing geodesic reaches $r_\infty$ in a finite time.

Using corollary~\ref{r-epsilon}, we only need to study the geodesic flow in a small neighborhood below $r_\infty$, as we know that it represents the point separating geodesics crossing the inner horizon or staying under it.

\section{The main result}
\label{sec:main-theorem}

We are finally ready to state the main result of this work, a theorem which allows one to find out whether all geodesics leaving at late times from the singularity are bound to cross the McVittie inner apparent horizon, or if there exists a time for which all geodesics leaving the singularity at subsequent times never reach the inner horizon and accumulate under it. In other words, we want to state here how the form of the Hubble parameter $H(t)$ can select the properties of the image of the inner horizon under the first-time projection function $\Phi$.

\begin{theorem}\label{oteorema}

Let there be the real-valued function $\Delta H (t) = H(t) - H_0$, and the constants $A = R (r_{\infty}) + r_{\infty} R\linha (r_{\infty})$, $B = R (r_{\infty}) \left( R\linha (r_{\infty}) - H_0 \right)$ and $t_i > 0$. If there exists $\delta > 0$ such that
\[
  F_+ (t_i, t) \equiv \int_{t_i}^t {e^{(B - \delta) u} e^{-A \int_{t_i}^u{\Delta H (s) \ud s}} \Delta H (u) \ud u}
\]
diverges as $t \to \infty$, then the image of $\mathcal{H}_-$ under $\Phi$ is unbounded.

Analogously, if there exists $\bar{\delta} > 0$ such that
\[
  F_-(t_i, t) \equiv \int_{t_i}^t {e^{(B + \bar{\delta}) u} e^{-A \int_{t_i}^u{\Delta H (s) \ud s}} \Delta H (u) \ud u}
\]
converges as $t \to \infty$, then the image of $\mathcal{H}_-$ under $\Phi$ is bounded.

\begin{proof}
Let us analyze the flow of Eq.~\eqref{hor-eq} near $r_{\infty}$, since we have already seen in Sec.~\ref{sec:rmin} that geodesics that cross $r_\infty$ are bound to reach the inner horizon in a finite time. Consider $0 < \epsilon <r_{\infty} - 2 m $. Let $z(t) = r_{\infty} - r(t)$, satisfying
\begin{gather}
  \dot{z} (t) = -X (t, r_{\infty} - z) \,,\nonumber
  \intertext{and}
  z (t_i) = z_0^{\epsilon} \,, \quad 0 < z_0^{\epsilon} < \epsilon \,.\nonumber
\end{gather}
Now, the crossing of $r_{\infty}$ by an ingoing geodesic is equivalent to a change of sign in $z(t)$. 

By the definition of the differential of $X$ (with respect to the second variable), at $r = r_{\infty}$, we may write
\[
  \dot{z} (t) = -X (t, r_{\infty})+  \ud X (t, r_{\infty}) z + \mathcal{O} (z)z \,.
\]
Then, there exists $\delta_{\epsilon} > 0$, continuous in $\epsilon$, such that $\lim_{\epsilon \to 0} \delta_{\epsilon} = 0$ and
\[
  \delta_{\epsilon}  \geq \mathcal{O} (z) \geq - \delta_{\epsilon} \,,
\]
for $0 < z < \epsilon$. We also define the curves $z^{\epsilon}_+ (t)$, $z^{\epsilon}_- (t)$, which are solutions of
\begin{gather}
 \dot{z}^{\epsilon}_{\pm} (t) = -X (t, r_{\infty}) + \left[ \ud X (t, r_{\infty}) \pm \delta_{\epsilon} \right] z^{\epsilon}_{\pm} \,,\label{Zequation}
\intertext{with initial condition}
 z_{\pm}^{\epsilon} (t_i) = z_0^{\epsilon} \,,\nonumber
\end{gather}
and verify $  z^{\epsilon}_-(t) \leq z(t) \leq z^{\epsilon}_+ (t)$ as long as $ \left| z^{\epsilon}_{\pm} (t) \right| \in \left[0, \epsilon \right)$. Therefore, if $z^{\epsilon}_- (t)$ is always positive, then $z(t)$ never changes sign and the geodesic $r(t)$ never crosses $r_\infty$. Conversely, if $z^{\epsilon}_+ (t)$ does change sign, then so does $z(t)$, and consequently, $r(t)$ crosses $r_\infty$.

The next step of the proof is to solve \eqref{Zequation}, which is a nonhomogeneous linear differential equation, formally solvable by the method of variation of constants. Explicitly,
\begin{align}
  \ud X (t, r_{\infty}) =& \left. \frac{\partial X}{\partial r} \right\vert_{r = r_{\infty}} \nonumber\\
  =&\, R\linha (r_{\infty}) \left[ r_{\infty} (H_0 + \Delta H (t)) - R (r_{\infty}) \right] \nonumber\\
  &+ R (r_{\infty}) \left[ H_0 + \Delta H (t) - R\linha (r_{\infty}) \right] \,,\nonumber
\end{align}
which, after using Eq.~\eqref{finfinito} and some rearrangement, gives us
\begin{align}
  \ud X (t, r_{\infty}) =&\, \left[ R\linha (r_{\infty}) r_{\infty} + R (r_{\infty}) \right] \Delta H (t) \nonumber\\
  & + R (r_{\infty}) \left[ H_0 - R\linha (r_{\infty}) \right] \nonumber\\
  =&\, A \Delta H (t) - B \,.
\label{dX}
\end{align}

We note that the last constant term in \eqref{dX} is
\[
-B \equiv  R (r_{\infty}) f_{\infty}\linha (r_{\infty}) = r_{\infty} H_0^2 - \frac{m}{(r_{\infty}^3 H_0)} \leq 0\,,
\]
as $f$ is positive between the singularity and the inner horizon. The case $f\linha_{\infty} (r_{\infty}) = 0$ corresponds to the extremal case where $f$ has two coincident real solutions and the spacetime asymptotes to an extremal Schwarzschild-de~Sitter, which has been discarded by hypothesis, since property \eqref{H-no-extreme} does not hold in this case. Thus,  $B > 0$. The constant 
\[
A \equiv R\linha (r_{\infty}) r_{\infty} + R (r_{\infty}) = \frac{m}{(r_{\infty}^3 H_0^2)} + r_{\infty} H_0
\]
is also strictly positive. 

The inhomogeneous term $X (t, r_{\infty})$ in \eqref{Zequation} can be written as
\begin{equation}
  X (t, r_{\infty}) = C \Delta H (t) \,, \label{Xinfty}
\end{equation}
where we define the positive constant
\[
C \equiv R (r_{\infty}) r_{\infty} =r_{\infty}^2 H_0 > 0 \,.
\]
Then, substituting Eqs.~\eqref{dX} and \eqref{Xinfty} into \eqref{Zequation}, we obtain
\begin{equation}
  \dot{z}^{\epsilon}_{\pm} (t) = \left[ A \Delta H (t) - B \pm \delta_{\epsilon} \right] z^{\epsilon}_{\pm} (t) - C \Delta H (t) \,,
\end{equation}
whose solutions are, with the initial conditions $z^{\epsilon}_{\pm} (t_i) = z_0^{\epsilon}$,
\begin{multline}
  z^{\epsilon}_{\pm}(t) = e^{\left(B\mp \delta_{\epsilon}\right) (t - t_i)} e^{A \int_{t_i}^t \Delta H (s) \ud s}\\
  \times \!\! \left[ z_0^{\epsilon} \! - \! C \!\! \int_{t_i}^t \!\! {e^{\left(B\mp \delta_{\epsilon}\right) (u - t_i)} e^{-A \int_{t_i}^u \Delta H (s) \ud s} \! \Delta H (u) \ud u} \right] \!.\label{Zsolution} 
\end{multline}

Next we need to find out under which conditions $ z^{\epsilon}_{\pm}$ change sign or not. Note that only the factor between brackets in \eqref{Zsolution} can be nonpositive, depending on the values taken by the integrals
\begin{align}
  &C \int_{t_i}^t e^{\left(B\mp \delta_{\epsilon}\right) (u - t_i)} e^{-A \int_{t_i}^u \Delta H(s)\ud s} \Delta H(u) \ud u \nonumber\\
  &\equiv C e^{\left(-B \pm \delta_{\epsilon}\right) t_i} F_{\pm}^{\epsilon} (t_i, t) \,.\nonumber
\end{align}
In particular, the convergence of $F_-$ or the divergence of $F_+$ in the $t \to \infty$ limit immediately tells us about the behavior of $z^\epsilon_\pm$. Given that $F_-(t_i, t) \geq F_+ (t_i, t)$, for all $\epsilon>0$ there are only three possible cases:

\begin{enumerate}[(a)]

 \item There exist $M (t_i) > 0$ and $\epsilon > 0$ such that $\lim_{t \to \infty} F_{-}^{\epsilon} (t_i,t) = M (t_i) $. \label{bounded}

 \item For all $\epsilon > 0$, $\lim_{t \to \infty} F_{+}^{\epsilon} (t_i, t) = \infty$. \label{unbounded}
 
 \item \label{dontknow} There exist $N(t_i) > 0$ and $\epsilon > 0$ so that $\lim_{t \to \infty} F_{+}^{\epsilon} (t_i,t) = N (t_i)$ but  $\lim_{t \to \infty} F_{-}^{\epsilon} (t_i,t) = \infty$ for all $\epsilon > 0$.

\end{enumerate}

Case (\ref{dontknow}) does not respect the hypothesis of the theorem; rather, it is the case in which the method presented here cannot be applied, and we will discuss it later. We can start then with case (\ref{bounded}) in which the term between brackets in \eqref{Zsolution} becomes
\[
  z_0^{\epsilon} - C e^{-\left( B + \delta_{\epsilon} \right) t_i} M (t_i) \,.
\]
However, as $\lim_{t_i \to \infty} C e^{-B t_i} M (t_i) = 0$, there exists $\tau > 0$ such that for all $t > \tau$, $z_0^{\epsilon} - C e^{-B t} M (t) > 0$. This means that after the instant $\tau$, the $z^{\epsilon}_-$ curves do not change sign anymore, and since they are a lower bound for $z$, neither do the curves with $t_i > \tau$. It follows that the ingoing geodesics $r_{t_0}(t)$, which reach $r_{\infty} - z_0^{\epsilon}$ at times equal to $\tau$ or later, do not cross the inner horizon at a finite coordinate time. As $t_0 < \tau$, this gives us an upper bound to the image of the inner horizon under the first-time projector as $\Phi (\mathcal{H}_-) \subset \left( 0, \tau \right]$.

In the case described by (\ref{unbounded}) instead, we see that there is no upper bound to $\Phi (\mathcal{H}_-)$. In fact, in this case there always exists a time $T_{t_i} > 0$ such that, for each $t_i > 0$,
\[
  F^\epsilon_+ (t_i, T_{t_i}) > \frac{z_0^{\epsilon} e^{\left( B - \delta_{\epsilon} \right) t_i}}{C} \,,
\]
which means that $z^{\epsilon}_+$ becomes negative for finite $t$, independently of the initial time $t_0$. Following the reasoning of case (\ref{bounded}), $z$ changes sign, and all ingoing geodesics eventually cross the inner horizon independently of the time at which they leave the initial singularity.
\end{proof}
\end{theorem}

Summarizing the results of the theorem, we have that the divergence (convergence) of the integral defining $F_+$ ($F_-$) allows us to find curves below (above) any geodesic around $r_\infty$ forcing them to cross (to stay below) the inner horizon. Crossing the inner horizon in a finite time independently of the starting point means, of course, that the geodesics that reach it leave the singularity at all times, while finding geodesics that never reach the inner horizon means that there exists a point in time where horizon-crossing geodesics accumulate.

\subsection{Limits of applicability}\label{subsec:dontknow}

As we said before, case (\ref{dontknow}) describes expansion functions for which we cannot immediately apply our method to find the asymptotes of the spacetime. The Hubble parameters that fall into this scenario are those that take the form $\Delta H(t) = e^{-B t} h(t)$, with $B$ the constant defined in the theorem, for which the conditions

\begin{enumerate}[(i)]

\item $\forall \, \epsilon > 0$, $\forall \, t_i \geq 0$, the integral $\int_{t_i}^{\infty} {e^{-\epsilon t} h(t)}\ud t$ converges, \label{fplusconv}

\item $\forall \, t_i \geq 0$, the integral $\int_{t_i}^{\infty} h(t) \ud t$ diverges, \label{fminusdiv}

\end{enumerate}
are both satisfied. These choices describe the ``blind spot'' of the method we presented. Although there is an infinite number of functions which may be constructed with these properties, the fine-tuning required by the exponential part means that such functions do not constitute a significant fraction of physically relevant cases, as we will see in an example in the next section.

\section{\texorpdfstring{Example: $\mathbf{\Lambda}$CDM expansion}{Example: ΛCDM expansion}}
\label{sec:example}

Let us illustrate the method described by theorem \ref{oteorema} by applying it to an example, which will also help show that most physically relevant expansion functions fall within the purview of the method. We choose the Hubble parameter used in $\Lambda$CDM models, where dark matter and dark energy are the main components of the energy budget, a good approximation to the Universe as we see it today, and also for the description of large-scale structure formation. In this case, $H$ can be described by \cite{ademir-plb-2010,lake-2011}
\begin{equation}\label{Hlake}
  H(t) = H_0 \coth \left( \frac{3}{2} H_0 t \right) \,.
\end{equation}

Keeping in mind that we want to analyze the asymptotes of the spacetime, we may expand \eqref{Hlake} at late times and rewrite it as
\[
  H(t) = H_0 + 2 H_0 e^{-3 H_0 t} + \mathcal{O} (e^{-6 H_0 t}) \,,
\]
which corresponds to $\Delta H = 2 H_0 e^{-3 H_0 t} \left[ 1 + \mathcal{O} (e^{-3 H_0 t}) \right]$. To prove that in this case the image of $\mathcal{H}_-$ is bounded, we need $F_- (t_i, t)$ to converge as per case (\ref{bounded}). On the other hand, if $F_+ (t_i, t)$ diverges, then we will have proved that the image of $\mathcal{H}_-$ is unbounded by falling into case (\ref{unbounded}) of the theorem.

With this form for $\Delta H$, the function $F_-$ is given by
\begin{multline}
  F_- (t_i, t) = \\ 
2H_0 \!\int_{t_i}^t \! \left[ 1 + \mathcal{O}(e^{-3 H_0 u}) \right] e^{(B - 3 H_0 + \delta) u -A \int_{t_1}^{u} \Delta H(s) \ud s} \ud u  \,.\nonumber
\end{multline}
Since the last factor is bounded between the two positive values
\[
 e^{-\frac{2A}{3} e^{-3 H_0 t_i \left[ 1 + \mathcal{O} (e^{-3 H_0 t_i}) \right]}}  \leq e^{-A \int_{t_1}^{u} \Delta H(s)\ud s} \leq 1 ,
\]
the convergence of $F_-$ is determined only by the integral
\[
\int_{t_i}^t e^{(B - 3 H_0 + \delta) u} \ud u \,.
\]
Therefore, $F_-$ will converge if there exists $\delta > 0$ such that $B - 3 H_0 + \delta < 0$. This is true if and only if
\[
  B - 3 H_0 < 0 \,.
\]
Following the same reasoning, we find that $F_+$ diverges if and only if
\[
  B - 3 H_0 > 0 \,.
\]

Inserting the definition for $B$, we notice that the relevant parameter for our discussion is given by
\begin{equation}\label{eta-def}
\eta \equiv \frac{B}{3H_0} - 1 = \frac{R (r_{\infty})}{3} \left[ \frac{R\linha (r_{\infty})}{H_0} - 1 \right] - 1 \,.
\end{equation}

Therefore, if $\eta <0$ then $F_-$ converges, and we have a limiting first time. If $\eta > 0$ then $F_+$ diverges, and all ingoing geodesics cross $r_-$. If $\eta = 0$, then $F_+$ converges and $F_-$ diverges, satisfying both properties (\ref{fplusconv}) and (\ref{fminusdiv}) in Sec.~\ref{subsec:dontknow}, and therefore leaving us with no knowledge about the ultimate fate of geodesics in the spacetime.

To calculate the values $\eta$ may assume, we use the fact that the constant $B$ can be explicitly calculated in terms of $r_{\infty}$. For McVittie metrics which asymptote to Schwarzschild-de~Sitter, the value of $r_{\infty}$ is given by \cite{stuchlik-1999}
\begin{equation}\label{rinf}
  r_{\infty} = \frac{2}{H_0 \sqrt{3}} \cos \left[ \frac{\pi}{3} + \frac{1}{3} \arccos \left( 3 \sqrt{3} m H_0 \right) \right] \,.
\end{equation}

\begin{figure}[!htp]
  \centering
  \includegraphics[width=0.45\textwidth]{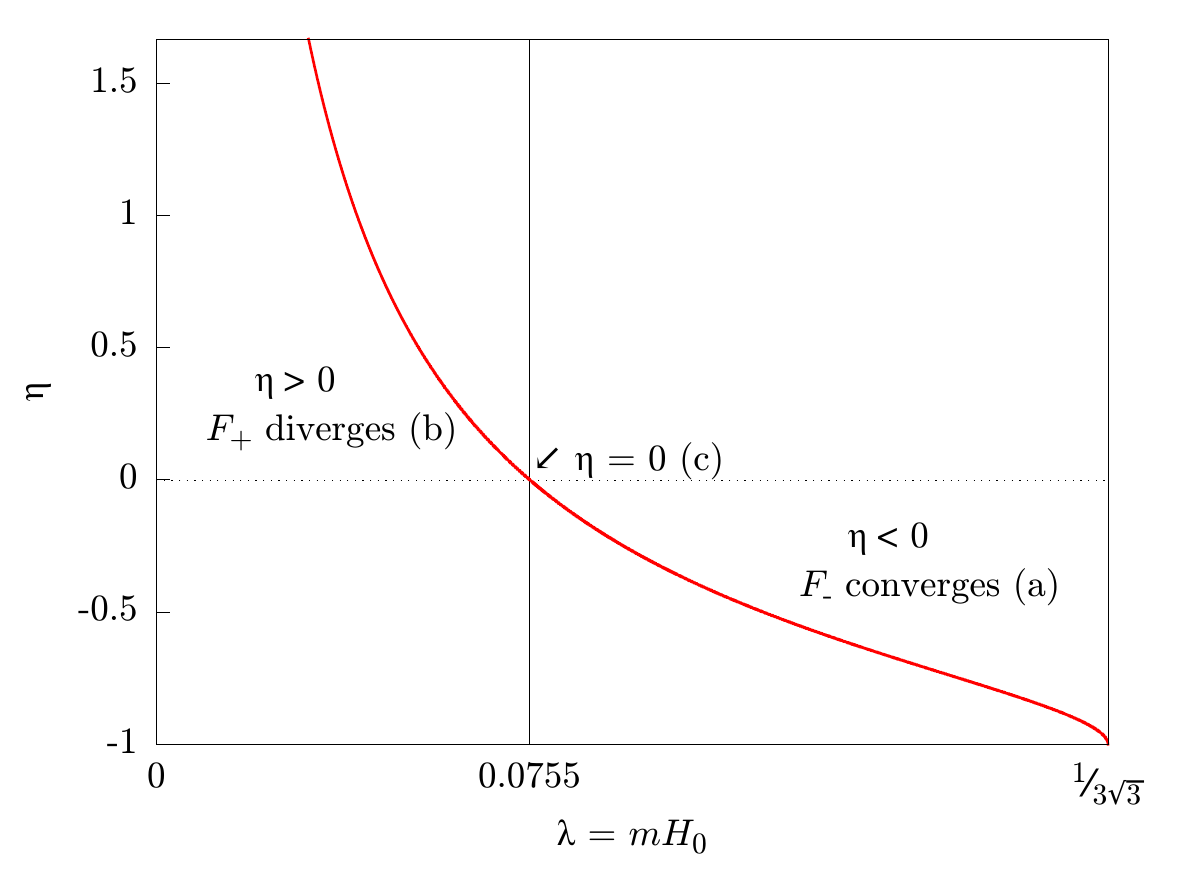}
  \caption{Convergence criterion applied to $\Lambda$CDM \eqref{Hlake}. Values of $\lambda = m H_0$ situated to the left of the root correspond to inner horizons which are always reached in finite time. Values situated to the right correspond to metrics with horizons that are only reachable by geodesics which start before a critical first time.\hspace*{\fill}}
  \label{eta-graph}
\end{figure}

It is easy to show that, once we insert \eqref{rinf} in \eqref{eta-def}, all dependence with respect to the free parameters $m$ and $H_0$ is expressed in terms of the product $\lambda = mH_0$; i.e., $\eta$ is constant along hyperbolas in the $m$--$H_0$ plane. Moreover, by noting that $m > 0$ and by making use of \eqref{H-no-extreme}, we find that the region of the parameter space available corresponds to $0 < \lambda < \frac{1}{3 \sqrt{3}}$. We plot the values taken by $\eta$ within this region in Fig.~\ref{eta-graph}.

In other words, in the region of the parameter space where $\eta < 0$ the causal diagram will be like the one in Fig.~\ref{diagrama-bound}, whereas where $\eta > 0$ the causal diagram will resemble the one from Fig.~\ref{causal-px}. In this model, $\eta$ has a root for $\lambda = 2 - \frac{10}{3 \sqrt{3}} \sim 0.0755 $. Only at this curve does our criterion fail to determine whether the image of $\mathcal{H}_-$ is bounded; therefore, in this case nothing can be said as to the final fate of geodesics.

\subsection{Consistency of various models}

As we hinted in Sec.~\ref{sec:rmin}, works that studied the asymptotics of the McVittie spacetime present two apparently incompatible results. Thanks to the analysis presented in this work, we can precisely define these two classes of models, knowing that their very existence and behavior are connected to the form of the expansion function $H(t)$. We can, in fact, say that there only exist two possibilities for the first-time projector of the inner horizon: $\Phi (\mathcal{H}_-)$ can be either bounded or unbounded. The former case was first studied in \cite{lake-2011}, where the $\Lambda$CDM model discussed above was considered with a value of $\lambda = \frac{958\,041}{6 \times 10^6} \sim 0.159$. In this case, the correct conclusion about the asymptotic behavior of the metric is that a white hole appears together with a black hole.

The other case, with an unbounded image for $\Phi(\mathcal{H}_-)$, has appeared before in the literature, and an example of it can be seen in \cite{kaloper-2010}, where, from the sketch of the causal structure the authors present, one can conclude that the choice of expansion function used satisfies case (\ref{unbounded}) of theorem \ref{oteorema} [or that it falls into case (\ref{dontknow}), but still gives an unbounded $\Phi(\mathcal{H}_-)$]. In reality, the authors of \cite{kaloper-2010} do not explicitly state their choice for the expansion function, except in a couple of simplifying examples where they assume a power-law expansion that gives $H(t) \propto \nicefrac{1}{t} + H_0$ (with $H_0$ taken to be a positive constant to compare to the results presented here). In any case, the analysis they present, which leads to the conclusion that the McVittie spacetime asymptotically tends to a black hole, applies only to the cases in question when $\Phi(\mathcal{H}_-)$ is unbounded. Therefore, provided that the adequate expansion function has been considered in the respective analysis, there is no friction between the discussion presented in \cite{lake-2011}, which claims that a white hole must be present together with a black hole in the asymptotics of McVittie spacetime, and other works where only the black hole is present.

\section{Conclusions}\label{sec:conclusions}

In this paper we developed a working method to determine the causal structure of McVittie metrics using the information contained in the Hubble parameter. One of two outcomes is possible: either geodesics leaving the singularity at all late times cross the inner horizon, corresponding to a case in which $\Phi (\mathcal{H}_-)$ is unbounded, or there exists a time for which any geodesic leaving the singularity after this upper bound never reaches the inner horizon, corresponding to a bounded $\Phi (\mathcal{H}_-)$. The causal structure that results from the first case is shown in Figs.~\ref{causal-px} and \ref{causal-x}, where the inner horizon at time infinity identifies with a black hole event horizon. The second case results in a causal structure as in Fig.~\ref{diagrama-bound}, where a bifurcating surface appears, splitting the boundary into a black hole in its future and a white hole in its past.

We showed that we can reduce the problem to the analysis of whether ingoing geodesics cross a fixed threshold given by the position of the inner horizon at future infinity. By using the linearized form of the geodesic equation we constructed majorant and minorant functions which can be treated analytically, and whose behavior near the threshold, due to the fact that they shepherd the geodesic between them into the same region, can be used to trace the solution of the full geodesic equation. The method is inconclusive only if the majorant crosses the threshold and the minorant does not, a situation which does not fix the behavior of the geodesic in between.

We applied this formalism, which is stated formally in theorem \ref{oteorema}, to analyze the causal structure of the McVittie metric when the expansion factor is given by a $\Lambda$CDM model. The simplicity of this example allowed us to cover the full spectrum of accessible values for the two independent parameters left in the metric, namely the black hole mass $m$ and the constant $H_0$, the asymptotic value of the Hubble parameter at time infinity. We found that both causal structures are possible, depending on the values of these parameters, and that the method we developed only fails at the curve given by $\eta = 0$ in the allowed portion of the two-dimensional parameter space for $m$ and $H_0$.

The aim of this work is not to discuss the meaning of the asymptotics of McVittie spacetime, especially since this has been done by various other authors (see, for instance, Refs.~\cite{kaloper-2010,lake-2011,faraoni-2012}). Here, the focus has been on finding a way to resolve the confusion present in the literature, namely, the appearance of apparently incompatible asymptotic behaviors of the McVittie spacetime, and to be able to distinguish these completely different physical setups that are generated from the same metric via different choices for the Hubble parameter. To clarify this important point, we used examples borrowed from the literature, in particular, identifying the two possible scenarios with the two cases studied in Refs.~\cite{kaloper-2010} and \cite{lake-2011} as shown in Figs.~\ref{causal-px} and \ref{diagrama-bound}, even though the results discussed here are fully general. With theorem \ref{oteorema} we proved that the form of the expansion function is the factor responsible for the structure of the boundaries in McVittie spacetime.

\begin{acknowledgments}

The authors wish to thank an anonymous referee for useful comments, J. C. A. Barata for helpful remarks during the first stages of the project, and K. Lake for pointing out a mistake in the first version of this paper and for an extended helpful discussion on the causal structure of the spacetime. D.C.G. also thanks B. Marin for programming tips. This work is supported by FAPESP and CNPq, Brazil.

\end{acknowledgments}

\bibliography{shortnames,referencias}

\end{document}